\documentclass[floatfix,aps,prl,twocolumn,groupedaddress]{revtex4}
\usepackage{graphicx}
\usepackage{amssymb}
\usepackage{color}
\usepackage{amsmath}
\usepackage{amsfonts}
\usepackage{hyperref}

\begin{document}

\title{Strongly correlated 2D quantum phases with cold polar molecules:\\
controlling the shape of the interaction potential}

\author{H.~P.\ B\"uchler$^{1,2}$}
\author{E.\ Demler$^{3}$}
\author{M.\ Lukin$^{3}$}
\author{A.\ Micheli$^{1,2}$}
\author{N.\ Prokof'ev$^{4}$}
\author{G.\ Pupillo$^{1,2}$}
\author{P.\ Zoller$^{1,2}$}
\affiliation{$^{1}$Institute for Theoretical Physics, University of Innsbruck, 6020
Innsbruck, Austria}
\affiliation{$^{2}$Institute for Quantum Optics and Quantum Information, 6020 Innsbruck,
Austria}
\affiliation{$^{3}$Harvard University, Department of Physics, Cambridge, USA}
\affiliation{$^{4}$BEC-INFM, Dipartimento di Fisica, Università di Trento, I-38050 Povo, Italy}

%\affiliation{$^{4}$Department of Physics, University of Massachusetts, Amherst, USA}
\date{\today }

\begin{abstract}
We discuss techniques to tune and shape the long-range part of the interaction
potentials in quantum gases of polar molecules by dressing rotational
excitations with static and microwave fields.  This provides a novel tool
towards engineering strongly correlated quantum phases in combination with low
dimensional trapping geometries. As an illustration, we discuss a 2D crystalline
phase, and a superfluid-crystal quantum phase transition.
\end{abstract}

\maketitle

% \pacs{03.75.Fi, 05.30.Jp, 32.80.Pj}

The outstanding feature of cold atomic and molecular quantum gases is the high
control and tunability of microscopic system parameters via external fields.
Prominent examples are the realization of low-dimensional trapping geometries,
and the tuning of the contact interaction via Feshbach resonances \cite{Control}.  In this
letter we extend this control to the \emph{shape} and the \emph{\ strength} of
interactions with the goal to generate new classes of potentials.  In
combination with low-dimensional trapping, these provide a framework for
realizing new many body quantum phases and phase transitions. We elaborate these
ideas in the context of polar molecules prepared in their electronic and
vibrational ground states~\cite{PolarMolecules}.

\begin{figure}[htb]
\begin{center}
\includegraphics[width=0.9\columnwidth]{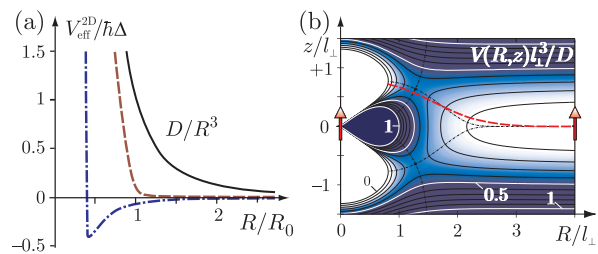}
\end{center}
\caption{(a) Effective potentials: dipole potential induced by
a static electric field $V_{\rm eff}^{\rm 2D}=
D/R^{3}$ (solid line); potential with a single microwave transition with $%
\Omega/\Delta=0.1$ (dashed line); attractive potential induced by an
additional microwave coupling to the states $|1,\pm1\rangle$ with
$\Delta_{\perp}=100\Delta$ and $\Omega_{\perp}/\Delta_{\perp}=0.3$
(dash-dotted line). (b) Contour plot of the potential
$V(\mathbf{r}) \: l_{\perp}^3/D$ (Eq.~\ref{microscopicpotential}) in the
$(R,z)$-plane. The instanton solution
is shown as dashed line. } \label{fig1}
\end{figure}

Cold polar molecules have attracted significant theoretical interest, in
particular in the context of dilute bosonic dipolar quantum gases
\cite{Collisions,Baranov05}. Special focus was on the appearence of
thermodynamic instabilities and roton softening for weakly interacting gases
\cite{Goral}.  Below we are specifically interested in polar molecules  in
\emph{ the strong interaction limit}, where the stability of the dipolar gas is
guaranteed by a confinement of the particles into a two-dimensional (2D) setup.
We show, that applying appropriately chosen static and/or microwave fields allow
us to design effective potentials $V_{\mathrm{eff}}^{\rm 2D}(\mathbf{R})$
between pairs of molecules (see Fig.~\ref{fig1}(a)). In turn, these potentials
give rise to interesting new many-body phenomena. As an illustration, we
consider the appearance of a crystalline phase, and an associated quantum
melting to a superfluid phase, for a 2D dipolar interaction
$V_{\mathrm{eff}}^{\rm 2D}(\mathbf{R})=D/R^{3}$.  The interaction strength is
characterized by the ratio between the interaction energy and the kinetic energy
$r_d=Dm/\hbar^{2}a$ with $a$ the average interparticle distance.  We determine
the transition point $r_{\mathrm{QM}}=18\pm4$ of this superfluid to solid
quantum phase transition via Path Integral Monte Carlo simulations. We find that
realistic experimental parameters of polar molecules allow for the realization
of this quantum phase, which has never been observed so far with cold atomic or
molecular gases.

We consider heteronuclear molecules prepared in their electronic and vibrational
ground states. We focus on bosonic molecules with a closed electronic shell
${}^{1}\Sigma (\nu =0)$, e.g. of the type SrO, RbCs or LiCs, as most relevant in
present experiments \cite{PolarMolecules}.  The Hamiltonian for a single
molecule $i$ is $H^{(i)}= \mathbf{p}_{i}^{2}/2m +
V_{\mathrm{trap}}(\mathbf{r}_{i}) +
H_{\mathrm{rot}}^{(i)}-\mathbf{d}_{i}\mathbf{E}(t).$ The first two terms are the
kinetic energy and the trapping potential for the center-of-mass motion of the
molecule of mass $m$. The term $H_{\rm rot}^{(i)}$ describes the internal low
energy excitations of the molecule, i.e., the rotational degree of freedom of
the molecular axis. This term is well described by a rigid rotor
$H_{\mathrm{rot}}^{(i)}=B\mathbf{J}_{i}^{2}$ with $B$ the rotational constant
(in the few to tens of GHz regime) and
$\mathbf{J}_{i}$ the dimensionless angular momentum. The rotational
states $|J,M\rangle $ with quantization axis $z$, and with
eigenenergies $BJ(J+1)$ are coupled by
a static or microwave field $\mathbf{E}$ via the electric dipole moment $%
\mathbf{d}_{i}$ (typically of the order of a few Debye). We assume a
setup where the molecules are confined to a 2D configuration in the
$x,y$-plane by a tight harmonic trapping potential $m\omega
_{\perp}^{2}z_{i}^{2}/2$ with $\omega_\perp=\hbar/ma_\perp^2$, as
provided by an optical potential. We remark that the different
dynamic polarizabilities parallel and perpendicular to the molecular
axis give rise to tensorshifts, which provide an  additional (small)
 state dependent potential for excited rotational
states~\cite{Friedrich95}.

The interaction of two polar molecules at distance
$r=|\mathbf{r}|=|\mathbf{r}_{1}-\mathbf{r}_{2}|$ is described by the Hamiltonian
$H=\sum_{i=1}^{2}H^{(i)}+V_{\mathrm{dd}}$ with the dipole-dipole interaction
$V_{\mathrm{dd}}=[\mathbf{d}_{1}
\mathbf{d}_{2}-3(\mathbf{d}_{1}\mathbf{n})(\mathbf{d}_2\mathbf{n})]/r^{3}$
and $\mathbf{n}=\mathbf{r}/r$ as unit vector \cite{Collisions}. In the absence
of external fields $\mathbf{E}$, the interaction of two molecules in their
rotational ground state is determined by the van der Waals attraction
$V_{\mathrm{vdW}}\sim -C_{6}/r^{6}$ with $C_{6}\approx
d^{4}/6B$, valid outside of the molecular core region $r>r_{\mathrm{rot}%
}\equiv (d^{2}/B)^{1/3}$. By dressing with an external field
$\mathbf{E}$, we induce and design long-range interactions. On a
formal level the derivation of the effective interaction proceeds in
two steps: (i) We derive a set of Born-Oppenheimer (BO) potentials
by first separating into center-of-mass and relative coordinates,
and diagonalizing the Hamiltonian
for the relative motion for fixed molecular positions $\mathbf{r}$, $H_{%
\mathrm{rel}}\equiv
\sum_{i=1,2}(H_{\mathrm{rot}}^{(i)}-\mathbf{d}_{i}\mathbf{E})+V_{\mathrm{dd}}$.
Within an adiabatic
approximation, the corresponding eigenvalues play the role of an effective $%
V_{\mathrm{eff}}^{\rm 3D}(\mathbf{r})$ interaction potential in a
given state manifold dressed by the external field~\cite{note1}.
(ii) We eliminate the motional degrees of freedom in the tightly
confined $z$ direction to obtain an effective 2D dynamics with
interaction $V_{\mathrm{eff}}^{\rm 2D}(\mathbf{R})$. In the
following we consider the cases of a static field and a microwave
field coupling the lowest rotor states.

A \emph{static electric field} applied perpendicular to the trap plane, $%
\mathbf{E}=E_{z}\mathbf{e}_{z}$, polarizes the rotational ground
state of each molecule, and induces finite dipole moments $\langle
\mathbf{d} _{i}\rangle \equiv \sqrt{D} \mathbf{e}_{z}$. These give
rise to a
long-range dipole-dipole interaction, $V_{\mathrm{eff}}^{\rm 3D}(\mathbf{r}%
)=D(r^{2}-3z^{2})/r^{5}$, where $D$ is tunable by the field strength
$E_{z}$. This expression of interacting dipoles aligned by the
external field is valid for distances larger than $r>r_{\star }$
with $r_{\star }$ defined by $C_{6}/r_{\star }^{6}\sim D/r_{\star
}^{3}$. Furthermore, for $r>r_{\mathrm{rot}}$ the BO approximation
is easily
fulfilled. The combination of the dipole-dipole interaction and the transverse {%
trapping} potential implies an interparticle potential
\begin{equation}
V(\mathbf{r})=\frac{D}{l_{\perp }^{3}}\left[ \frac{%
l_{\perp }^{3}}{r^{3}}-3\frac{z^{2}l_{\perp }^{3}}{r^{5}}+\frac{z^{2}}{%
2l_{\perp }^{2}}\right]   \label{microscopicpotential}
\end{equation}%
with $l_{\perp }=(Dm/2\hbar ^{2}a_{\perp })^{1/5}a_{\perp }$  (see Fig.~\ref%
{fig1}(b)), and where we require $l_{\perp }$ $\gg $ \textrm{max}$(r_{\star
},r_{\mathrm{rot}})$. The above potential has a saddle point at $%
R_{s}\approx 1.28l_{\perp }$ and $z_{s}\approx \pm 0.64l_{\perp }$
with the potential height $V_{s}\approx 0.34D/l_{\perp }^{3}$. As a
consequence, the short distance regime, $r<l_{\perp }$, is separated
by a potential barrier from the large distance regime, $r>l_{\perp
}$. Within a semi-classical
approximation, %\cite{Coleman77}, 
the tunneling rate through this barrier takes the form
$\Gamma =\omega _{p}\exp [-c(Dm/2\hbar ^{2}a_{\perp })^{2/5}]$ with
$c\approx 5.86$ and  $\omega _{p}\sim \sqrt{D/ma^{5}}$  the
\textquotedblleft attempt frequency\textquotedblright.
Therefore, in the strongly interacting limit with a tight confinement
the tunneling
rate is exponentially suppressed, and the probability to find two
particles at a distance $r<l_{\perp}$ is negligible, and  guarantees the
stability of an ensemble of polar molecules. %~\cite{note2}. 
Then, the system is completely determined by the interaction
potential at large distances $r\gg
l_{\perp }$, which is in particular also much larger than the short distances
scales $r_{\star },r_{\mathrm{rot}}$. Thus we can reduce the interaction to an
effective 2D potential by integrating out the fast transverse motion
as
\begin{equation}
V_{\mathrm{eff}}^{\rm 2D}(\mathbf{R})=\int dz_{1}\int dz_{2}|\psi
_{\perp }(z_{1})|^{2}|\psi _{\perp
}(z_{2})|^{2}V_{\mathrm{eff}}^{\rm 3D}(\mathbf{r}), \label{trap2D}
\end{equation}%
where $\psi _{\perp }(z_i)=\exp (-z_i^{2}/2a_{\perp }^{2})/(\pi
a_{\perp }^{2})^{1/4}$ is the ground state harmonic oscillator
wave-function in the tightly confined $z$-direction, and is valid for distance
$R> (D/\hbar \omega_{\perp})^{1/3}$ 
(note, that $l_{\perp} \sim (D/\hbar \omega_{\perp})^{1/3}$ for realistic
parameters). Therefore, the effective 2D interaction reduces to
$V_{\mathrm{eff}%
}^{\rm 2D}\sim {D/R^{3}}$ for large separations $R\gg
l_{\perp }$.

\emph{Microwave fields} can drive the transition of a molecule from the ground
to the first excited rotational state. We denote the corresponding Rabi
frequency by $\Omega$ and the detuning by $\Delta $. In the weak driving limit
$\Omega <\Delta $ the molecules are essentially in their rotational ground state
with AC Stark shifted energy and a small admixture from the excited state. For
two molecules approaching each other and  blue detuning $ \Delta >0$ (cf. Fig.  2), the
microwave field will be resonant at distance $R_{0}$ with
$d^{2}/R_{0}^{3}\sim \hbar\Delta$, i.e., the ground and first excited rotational states 
will be strongly mixed.
This gives rise to an effective BO potential, which for $r<R_{0}$ exhibits a  strong
repulsion inherited from the  $\sim 1/r^{3}$
excited state dipole-dipole interactions; Similar ideas have been
discussed as ``blue shield'' in the context of alkali
atoms~\cite{Weiner99}.

%which for $r>R_{0}$ reflects the (weak)
%interactions between ground state molecules but for $r<R_{0}$ has a
%strongly repulsive potential inherited from the $\sim 1/r^{3}$
%excited state dipole-dipole interactions; Similar ideas have been
%discussed as ``blue shield'' in the context of alkali
%atoms~\cite{Weiner99}.

\begin{figure}[htb]
\begin{center}
\includegraphics[width=\columnwidth]{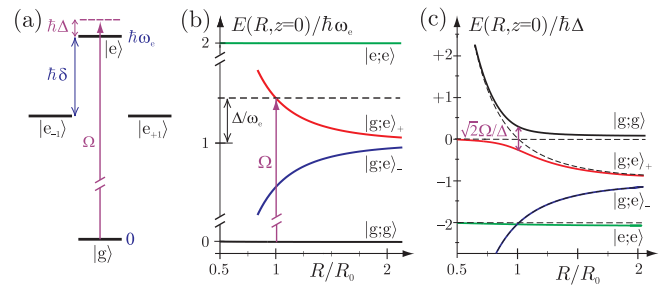}
\end{center}
\caption{(a) Energy levels of the rotor in a weak electric field.
The microwave transition with detuning $\Delta$ and Rabi frequency
$\Omega$ is shown as the arrow. (b) BO potentials for the internal
states for $\Omega=0$, where
$|g;e\rangle_\pm\equiv(|g;e\rangle\pm|e;g\rangle)/\sqrt{2}$ denote
the gerade/ungerade first excited states. (c) Avoided level crossing
due to the microwave coupling with the effective BO potential $V_{\rm eff}^{\rm 3D}$
given by the highest eigenenergy state.}\label{fig2}
\end{figure}

As an example, we consider a microwave field with polarization along the
$z$-axis, driving the transition between the ground state and the first excited
state with $M=0$. Excitations into higher states are suppressed due to the
anharmonicity of the spectrum, and spontaneous emission is irrelevant in the
microwave regime. The interaction Hamiltonian also couples the excited state
$|1,0\rangle$ to the  states $|1,\pm 1\rangle$ for  non-vanishing $z$ (note that such a
coupling can also be induced by the tensor shift). In the following, we suppress this
coupling by introducing a (weak) static electric field $E_{z} \ll B/d$ lifting the
degeneracy of the $J=1$ manifold by $\delta \gg \Delta$, see
Fig.~\ref{fig2}(a-b), and we focus on distances $r>r_{\delta}=
(d^2/\hbar\delta)^{1/3}$. For each
polar molecule, the system reduces to a two level system with the
ground state $|g\rangle$ and the first excited state $|e\rangle$
($M=0$) dressed by the  electric field, and energy difference
$\hbar \omega_{e}$.  In a rotating frame the Hamiltonian $H_{\rm
rel}$ expressed in the basis $\{|g;g\rangle, |g;e\rangle,
|e;g\rangle, |e;e\rangle\}$ becomes
\begin{equation}
  H_{\rm rel} = \hbar
\left[
\begin{array}{cccc}
  D_{g} \nu({\bf r}) &  \Omega&  \Omega & 0\\
   \Omega &- \Delta&  D \nu({\bf r})  &  \Omega \\
 \Omega & D \nu({\bf r})& -  \Delta &  \Omega \\
0&  \Omega &  \Omega & D_{e} \nu({\bf r})-2  \Delta
\end{array}
\right],
\end{equation}
where $D_{g}= |\langle g| {\bf d}_{i} |g\rangle|^2$ and $D_{e}=
|\langle e| {\bf d}_{i} |e\rangle|^2$ denote the weak dipole
interaction induced by the static field $E_{z}$, while $D\approx d^2/3$ 
accounts for the dipole transiton element between the
ground and excited state, and $\nu({\bf r}) = (r^2-3z^2)/\hbar r^5$
is the shape of the dipole interaction.  The effective BO potential
$V_{\rm eff}^{\rm 3D}({\bf r})$ derives now from the eigenenergy of
the state adiabatically connecting to  $|g,g\rangle$ for
$r\rightarrow \infty$, and is given by the highest energy level, see
Fig.~\ref{fig2}(c).  We find that for a detuning  $\Delta$ with
$R_{0}= (D/\hbar\Delta)^{1/3} \gg l_{\perp}\gtrsim r_{\delta}$ the
combination of the transverse trapping potential with $V_{\rm
eff}^{\rm 3D}$ provides again a large tunneling barrier separating
the short distance regime from the long distance regime. We
emphasize that the existence of this barrier makes the system stable
-- in contrast to the incomplete shielding discussed previously in
the context of ``blue shielding'' with lasers~\cite{Weiner99}.
Then, in
analogy to the discussion of the static electric field, we obtain
the effective 2D potential via integrating out the transverse
motion; the effective potential $V_{\rm eff}^{\rm 2D}$ is shown in
Fig.~\ref{fig1}(a). The distance $R_{0}$ separates a weakly
interacting regime at large distances $R>R_{0}$ with $V^{\rm
2D}_{\rm eff} \sim \left[ (2\Omega^2/\Delta^2) D+ D_{g}\right] /
R^3$ from a strongly repulsive regime with $V^{\rm 2D}_{\rm eff}\sim
D/R^3$ on distances $l_{\perp}< R< R_{0}$.

The BO approximation above is valid as long as the passage through
the level crossing at $R_0$ is adiabatic. For a realistic setup with
average interparticle distance $a\sim R_{0}$, the average particle
velocity can be estimated as $\hbar/R_{0} m$. Then, the Landau-Zener
probability for a diabatic crossing reduces to
$P_{\mathrm{\scriptscriptstyle LZ}} = \exp\left[ -2 \pi (D m
\Omega^2/\hbar^{2} R_{0}\Delta^{2})\right] , $ and the BO
approximation is valid for $\Delta^{2}/\Omega^{2} < D m/\hbar^{2}
R_{0}$. This condition competes with the weak coupling constraint $
\Omega^{2}/\Delta^{2} <1$. However, for $\Omega/\Delta\sim0.1$, both
conditions can be satisfied for strong dipole interactions with
$r_{d}>100$.

The use of additional microwave fields coupling to the $|1,\pm1\rangle$-states
with detuning $\Delta_{\perp}\gg \Delta$ and Rabi frequency $\Omega_{\perp}$
allows for further shaping  the effective potentials. One such
example is shown in Fig.1(a), where $V_{\mathrm{eff} }^{\rm 2D}$ in the long
range part $R>R_{0}$ becomes attractive and allows for the existence of
bi-molecular bound states.

\begin{figure}[htb]
\begin{center}
\includegraphics[width=\columnwidth]{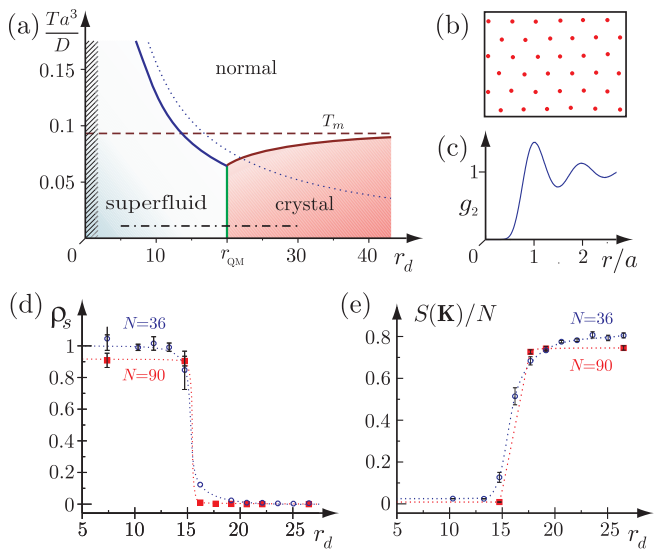}
\end{center}
\caption{(a) Tentative phase diagram in the $T-r_{d}$ plane:
crystalline phase for interactions $r_{d}>r_{\rm QM}$ and temperatures below the
classical melting temperature $T_{m}$ (dashed line). The superfluid phase
appears below the upper bound $T<\protect\pi \hbar^{2}n/2 m $ (dotted line).
The quantum melting transition is studied at fixed temperature $T=0.014 D/a^{3}$
with interactions $r_{d}=5-30$ (dash-dotted line). The crossover to the unstable
regime for small replusion and finite confinement $\omega_{\perp}$ is indicated
(hatched region). (b) PIMC-snapshot of the mean particle positions in the
crystalline phase for $N=36$ at $r_{d}\approx26.5$. (c) Density-density
(angle-averaged) correlation function $g_{2}(r)$, for $N=36$ at $r_{d}\sim11.8$.
(d) Superfluid density $\rho_{s}$ and (e) static structure factor $S({\bf K})/N$
as a function of $r_{d}$, for $N=36$ (circles) and $N=90$ (squares).}
\label{fig3}
\end{figure}

Finally, we can extend the above discussion 
to a gas of cold polar molecules
confined into 2D with temperature $T<\hbar \omega_{\perp}$, and obtain the {\it
many body Hamiltonian}~\cite{note3}
\begin{equation}
 H = \sum_{i} \frac{{\bf p}_{i}^{2}}{2 m} + \sum_{i < j}
 V_{\rm eff}^{\rm 2D}({\bf R}_{i}- {\bf R}_{j}). \label{hamilton}
\end{equation}
The first term accounts for the kinetic energy within the $x$,$y$-plane, while
the second term  denotes the effective interaction potential with the strong
repulsion $\sim D/R^3$ on short distances $r<R_{0}$, see above. The validity of
this effective Hamiltonian  requires that tunneling events through the barrier
of the interparticle potential are suppressed, i.e., strong interactions and tight confinement with $D
m/\hbar^2 a_{\perp}\gg 1$.  In turn, for decreasing interactions, tunneling
through the barrier takes place and three body recombinations become relevant
driving a crossover into a potentially unstable regime (note, that the approach towards
this intermediate region from the weakly interacting side has been previously
discussed \cite{Goral}).  The Hamiltonian Eq.~(\ref{hamilton}) gives rise to
novel quantum phenomena, which have not been accessed so far in the context of
cold atoms/molecules.  As an illustration, we focus on the  interaction $V_{\rm
eff}^{\rm 2D}({\bf R})= D/R^3$  with the dimensionless parameter $r_{d}=
Dm/\hbar^2 a$ and the particle density $n=1/a^2$. 
In the following, we present a tentative phase diagram of the system, see
Fig.~\ref{fig3}, and show that for realistic experimental parameters the polar
molecules can be driven from the superfluid into the crystalline phase.

In the limit of strong interactions  $r_{d}=Dm/\hbar^{2} a \gg1$
the polar molecules are in a crystalline phase for temperatures $T< T_{m}$ with 
$ T_{m}\approx0.09 D/a^{3}$ \cite{Kalia81}. The
configuration with minimal energy is a triangular lattice with spacing $a_{%
\mathrm{\scriptscriptstyle L}} = (4/3)^{1/4} a$. The excitations are two linear
sound modes with characteristic phonon
frequency $\omega_{p}= \sqrt{D/ a^{5} m}$.
The static structure factor
$S$ diverges at a reciprocal lattice vector $\mathbf{K}$, and $S(\mathbf{K}%
)/N$ acts as an order parameter for the crystalline phase.
In the opposite limit of weak interactions $r_{d}<1$, the ground
state is a superfluid (SF) with a finite (quasi) condensate. The SF
is characterized by
a superfluid fraction $\rho_{s}(T)$, which depends on temperature $T$, with $%
\rho_{s}(T=0)=1$. For finite temperature, a Berezinskii--Kosterlitz--Thouless transition %\cite{KosterlitzThoulessJPhysC6-1181-1973}
towards a normal fluid is expected to occur at
$T_{\mathrm{\scriptscriptstyle KT}}= \pi\rho_{s}\hbar^{2}n/2 m$.
In turn, very little is known on the intermediate strongly interacting regime
with $r_{d}\gtrsim1$; see \cite{Arkhipov05} for a discussion in 1D. Here, we
focus on this intermediate regime and determine the critical interaction
strength $r_{\mathrm{ \scriptscriptstyle QM}}$ for the quantum melting
transition. We use a recently developed PIMC-code based on the Worm algorithm
\cite{Boninsegni06} . In Fig.~\ref{fig3}(d-e), the order parameters $\rho_{s}$
and $S(\mathbf{K})/N $ are shown at a small temperature $T= 0.014 D/a^{3}$ for
different interaction strengths $r_{d}$ and particle numbers $N=36,90$. We find
that $\rho_{s}$ exhibits a sudden drop to zero for $r_{d}\approx 15$, while at
the same position $S(\mathbf{K})$ strongly increases. In addition, we observe
that 
in a few occasions $\rho_s$ suddenly jumped
from 0 to 1, and then returned to 0, in the interval
$r_d\approx 15-20$, which suggests a competition between the
superfluid and crystalline phases. These results
indicate a superfluid to crystal phase transition at $r_{\mathrm{\scriptscriptstyle QM%
}}= 18\pm4$. The step-like behavior of $\rho_{s}$ and $S({\bf K})/N$ is consistent with a first order phase transition. 
For $r_{d}>r_{\mathrm{\scriptscriptstyle QM}}$ the crystalline structure is
triangular, in agreement with the discussion above, see Fig.~\ref{fig3}(b). 
Note, that the superfluid with $r_{d}\sim 1$ is strongly interacting,
% and shows
%several differences compared to a Bose-Einstein condensate of cold atoms.
in particular, the density-density correlation function is quenched on lengths 
$R<a$, see Fig.~\ref{fig3}(c).
Moreover, the (quasi) condensate involves a fraction of the total density, and therefore small coherence peaks in a
time of flight experiment are expected.
However, the detection of vortices can be used as a definitive signature of
superfluidity, while Bragg scattering with optical light allows for 
probing   the crystalline phase.

While many of the molecular species studied at present in the lab
are candidates for observing the above phenomena, we note that e.g.
$\mathrm{SrO}$ has a large dipole moment $d=8.9D$ and allows optical
trapping with a red detuned laser with wave length $\lambda\sim1{\rm
{\mu}m}$. Then, for a trapping potential with $a_{\perp}\sim 50
\mathrm{nm}$ and interparticle distance $a \sim 300 \mathrm{nm}$, we obtain
$r_{d}\sim410 \gg r_{\mathrm{%
\scriptscriptstyle QM}}$, and the
tunneling rate is suppressed with $d^{2} m/ \hbar^{2} a_{\perp}
\sim 2500$. Since the classical melting occurs at
$T_{m}(r_{d}=400)\sim2 \mu\mathrm{K}$, this molecule is a candidate
to reach the quantum regime.

In conclusion, we have shown that cold polar molecules allow to
extend the control of interactions in cold gases to the
\textit{shape} of the inter-particle potential, which opens a route
towards generation of (novel) strongly correlated quantum phases in
low dimensional trapping geometries. In particular, 
we have
discussed the appearance and properties of a crystalline phase, and
an associated quantum phase transition to a dipolar quantum gas.
Besides the fundamental interest of observing crystal phases with
neutral particles in atomic and molecular physics, the properties of
dipolar molecular crystals might lead to interesting applications:
these crystals appear in the high density limit, i.e. for small
interparticle separation, which could potentially be smaller than
optical lattice structures. At the same time, crystal phases are
expected to be stable in the sense that (possibly damaging) short
range collisions are suppressed.

Work in Innsbruck was supported by Austrian Science Foundation, the
European Union under contracts FP6-013501-OLAQUI,
MRTN-CT-2003-505089 and IST-15714, and the Institute for Quantum
Information; N.~P. was supported by NSF grant PHY-0426881;


\begin{thebibliography}{99}

%\bibitem{Review}{For a review, see: J. I. Cirac and P. Zoller, Physics Today \textbf{57}, 38
%(2004); F.~Chevy and C.~Salomon, Physics World~\textbf{18}, 43
%(2005).}


\bibitem{Control} {Nature insight: ultracold matter, Nature \textbf{416}, 205 (2002)}

\bibitem{PolarMolecules}{See e.g.~Special Issue on ultracold polar molecules, Eur.~Phys.~J.~D~\textbf{31}, 149-445
(2004); J.~M.~Sage, \textit{et al.}, Phys. Rev. Lett.~\textbf{94},
203001 (2005);
 T.~Rieger, \textit{et al.}, {\it ibid.} \textbf{95}, 173002 (2005);
 D. Wang, \textit{et al.}, {\it ibid.} \textbf{93}, 243005 (2004);
 S.~Y.~T.~van~de~Meerakker, \textit{et al.}, {\it ibid.} \textbf{\ 94}, 023004 (2005);
  S.~D.~Kraft, \textit{et al.}, arXiv:physics/0605019.}
% Doyle; DeMille; Rempe; Goudl-Stwalley (KRb), WeideMueller; Maijer (bad!, since OH)

\bibitem{Collisions}{C. Ticknor and J.~L. Bohn, Phys. Rev. A \textbf{72}, 032717 (2005);
R.~V.~Krems, Phys. Rev. Lett. \textbf{96}, 123202 (2006).}


\bibitem{Baranov05}{K.~Goral, L.~Santos, and M.~Lewenstein, Phys. Rev. Lett. 88, 170406
(2002); M.~A.~Baranov, K.~Osterloh, and M.~Lewenstein,
\textit{ibid.}~\textbf{94}, 070404 (2005); E. H. Rezayi, N. Read,
and N. R. Cooper, \textit{ibid.}~\textbf{95}, 160404 (2005);
A.~Micheli, G. K. Brennen, and P.~Zoller, Nature Physics~\textbf{2},
341 (2006).}


\bibitem{Goral} {K.~Goral, K. Rzazewski, and T.~Pfau, Phys.
Rev. A~{\bf 61}, 051601(R) (2000); L.~ Santos, G. V. Shlyapnikov, and
M. Lewenstein, Phys. Rev. Lett.~\textbf{90}, 250403 (2003); J.
Stuhler, \textit{et al.}, \textit{ibid.}~\textbf{95}, 150406
(2005).}


%\bibitem{Grimes79}{C.~C.~Grimes and G.~Adams,  Phys.~Rev.~Lett.~\textbf{42},
%795 (1979).}

%E.~Y.~Andrei, ``Two Dimensional Electron Systems on Helium and Other
%Cryogenic Substrates'', Academic Press, New York (1991).

\bibitem{Friedrich95} {B.~Friedrich and D.~Herschbach,  Phys. Rev. Lett.
\textbf{74}, 4623 (1995); S. Kotochigova and E. Tiesinga,  Phys.
Rev. A \textbf{73}, 041405(R) (2006).}

%\bibitem{Coleman77}{S. Coleman, Phys. Rev. D~\textbf{15}, 2929
%(1977).}

\bibitem{Weiner99}{J.~Weiner, \textit{et al.}, Rev. Mod. Phys. {\bf 71}, 1 (1999).}
%  John Weiner, Vanderlei S. Bagnato and Sergio Zilio, Paul S. Julienne
% Experiments and theory in cold and ultracold collisions

\bibitem{Kalia81}{R. K. Kalia and P. Vashishta, J. Phys.
C~\textbf{14}, 643 (1981).}

%\bibitem{Strandburg88}{K. J. Strandburg, Rev. Mod.
%Phys.~\textbf{60}, 161 (1988).}

\bibitem{Arkhipov05}{A. S. Arkhipov, \textit{et al.}, JETP Lett.~{\bf 82}, 39
(2005).} % Astrkharchik, Belikov, Lozovik

\bibitem{Boninsegni06}{M. Boninsegni, N. Prokof'ev, and B. Svistunov,
Phys. Rev. Lett.~\textbf{96}, 070601 (2006).}

%\bibitem{Stuhler06}{}


%\bibitem{Bonsall77}{L. Bonsall and A. A. Maraududin, Phys. Rev.
%B~\textbf{15}, 1959 (1977)}

%\bibitem{Tanatar89}{B. Tanatar and D. M. Ceperley, Phys. Rev.
%B~\textbf{39}, 5005 (1988).}

%\bibitem{Guido}{G. Pupillo, \textit{et al.}, to be
%published.}

\bibitem{note1}{For
a sinusoidal $\mathbf{E}(t)$ a diagonalization precludes a
transformation to a time-independent Floquet Hamiltonian.}

%\bibitem{note2}{For decreasing interaction strength the tunneling through the barrier
%increases, and the system enters an unstable regime.}

%, which separates the strongly interacting stable domain from a
%weakly interacting dipolar gas.

\bibitem{note3}{Three body
interactions can be important for particles approaching each other
on distances $|{\bf R}_{i}-{\bf R}_{j}|\sim R_{0}$, but these corrections are dropped in the following.}


%Evidence for a Liquid-to-Crystal Phase Transition in a Classical,
%Two-Dimensional Sheet of Electrons
%    C. C. Grimes and G. Adams
%    Bell Laboratories, Murray Hill, New Jersey 07974
%Received 17 January 1979
%Experimental evidence is presented for an electron-liquid to
%electron-crystal phase transition in a sheet of electrons on a
%liquid-He surface. The phase transition has been studied for
%electron areal densities from 3×108 cm-2 to 9×108 cm-2 and has
%yielded melting temperatures between 0.35 and 0.65 K. The phase
%transition occurs at ?=137±15, where ? is a measure of the ratio of
%potential energy to kinetic energy per electron.



\end{thebibliography}
\end{document}